%% file: arxiv-v2.tex
\documentclass[twocolumn,superscriptaddress,pre]{revtex4-2}

\include{preamble}

\usepackage[english]{babel}
\usepackage[utf8]{inputenc}
\usepackage[T1]{fontenc}
\usepackage{CJK}
\usepackage{microtype}

\begin{document}

\title{Learning nonequilibrium control forces to characterize dynamical phase transitions}

\author{Jiawei Yan (闫嘉伟)}
\affiliation{Department of Chemistry, Stanford University, Stanford, CA 94305, USA}
\author{Hugo Touchette}
\affiliation{Department of Mathematical Sciences, Stellenbosch University, Stellenbosch 7600, South Africa}
\author{Grant M. Rotskoff}
\affiliation{Department of Chemistry, Stanford University, Stanford, CA 94305, USA}
\email[To whom correspondence should be addressed:]{rotskoff@stanford.edu}
\date{\today}

\begin{abstract}
Sampling the collective, dynamical fluctuations that lead to nonequilibrium pattern formation requires probing rare regions of trajectory space. Recent approaches to this problem, based on importance sampling, cloning, and spectral approximations, have yielded significant insight into nonequilibrium systems, but tend to scale poorly with the size of the system, especially near dynamical phase transitions. Here we propose a machine learning algorithm that samples rare trajectories and estimates the associated large deviation functions using a many-body control force by leveraging the flexible function representation provided by deep neural networks, importance sampling in trajectory space, and stochastic optimal control theory. We show that this approach scales to hundreds of interacting particles and remains robust at dynamical phase transitions.
\end{abstract}

\begin{CJK*}{UTF8}{}
\CJKfamily{gbsn}
\maketitle
\end{CJK*}

\section{Introduction}

Large deviation techniques have been used recently to gain physical insight into the steady state and fluctuations of a diverse set of systems driven away from equilibrium, including diffusive and colloidal systems~\cite{zon2004a,ciliberto2010,ciliberto2017}, glassy dynamics~\cite{merolle2005,garrahan_dynamical_2007,hedges_dynamic_2009,chandler2010}, interacting particle systems driven by external reservoirs~\cite{derrida2007,bertini2007,bertini2015b}, and active matter~\cite{cagnetta2017,grandpre2018,whitelam2018e,keta2021}. 
Fluctuations of dynamical quantities, such as currents and kinetic activities, provide information about complex pattern formation and phase behavior that can emerge in these systems when detailed balance is broken. The study of nonequilibrium fluctuations has also led to the discovery of fundamental results, such as the fluctuation relation~\cite{crooks_entropy_1999,lebowitz_gallavotti_1999,harris2007}, which encodes symmetries in the distribution of the entropy production, and, more recently, the thermodynamic uncertainty relation~\cite{pietzonka_universal_2016,barato_dispersion_2015,gingrich_dissipation_2016}, which connects current fluctuations to dissipation.

The likelihood of fluctuations is described in large deviation theory by functions playing the role of nonequilibrium potentials that are notoriously difficult to compute for complex and high-dimensional systems. While analytical treatment is possible for some systems~\cite{derrida_exact_1998,lazarescu_exact_2015,bodineau_current_2006}, we must generally estimate these functions numerically. Many algorithms have been proposed for this purpose, based either on spectral methods or on sampling rare trajectories, using a combination of importance sampling~\cite{bucklew2004,nemoto2016,ray2018,ferre2018}, cloning~\cite{grassberger2002,giardina2006,lecomte2007a,ray2020}, and reinforcement learning~\cite{das_variational_2019, rose_reinforcement_2021, das_reinforcement_2021}. Good results are reported with most methods, although it remains challenging to obtain good convergence in systems with many degrees of freedom, especially when probing fluctuations near phase transitions~\cite{nemoto2016}. 

In this paper, we present an algorithm that combines control theory, importance sampling, and, crucially, the robust and flexible function representations offered by neural networks to calculate large deviation functions. The algorithm uses recent developments in machine learning approaches to PDEs~\cite{e_deep_2018,rotskoff_active_2021} and estimates large deviation functions by adaptively constructing a many-body control force that drives a nonequilibrium system of interest in an optimal way towards a given dynamical fluctuation. Unlike other methods that construct a control force, our approach is based on a direct stochastic optimization of a cost functional for trajectories, in which gradients are computed through the dynamics or via an adjoint stochastic dynamics, which is robust over long trajectories~\cite{li_scalable_2020}. 

We illustrate our algorithm with two stochastic models: a simple diffusion showing a dynamical phase transition in the low-noise limit and a model of active Brownian particles driven by pair interactions and an alignment force. The results for both show that our approach is robust near dynamical phase transitions and efficiently scales to large systems of interacting particles, which are difficult to treat with spectral methods or cloning algorithms. For the active Brownian particle model, we are able for instance to estimate large deviation functions for systems of up to 200 particles, which is unreachable for cloning without substantial computational power.  Our algorithm requires fewer parallel replicas than cloning algorithms, uses much less memory by relying on single trajectories, and converges faster, as we demonstrate with the simple diffusion model.

\section{Model and large deviations}

We consider systems described by a stochastic differential equation (SDE) having the general form
\begin{equation}
\label{eq:sde}
    d\Xb_t = b(\Xb_t) dt + \sigma d\Wb_t,
\end{equation}
where $\Xb_t\in\RR^d$ is the state of the system, $b:\RR^d \to \RR^d$ is the drift function, and $\Wb_t$ is a Wiener process acting as a noise source, which is multiplied by the noise matrix $\sigma$. This model captures the diffusive dynamics of many physical systems, as the drift or ``force'' $b(\xb)$ can include the gradient of a many-body potential energy describing the interactions among a large number of particles, in addition to non-conservative and hence nonequilibrium external forces. We assume that the drift and the noise source are such that $\Xb_t$ is ergodic, so that it has a unique probability stationary density, reached from any initial distribution in the long-time limit. For simplicity, we also assume that $\sigma$ is independent of $\xb$ and that the corresponding diffusion tensor $D= \sigma\sigma^T$ is invertible. 

Given the dynamics for $\Xb_t$, we are interested in finding the distribution of time-integrated or ``dynamical'' observables having the form
\begin{equation}
 A_T = \frac1T \int_0^T f(\Xb_t) dt + \frac1T \int_0^T g(\Xb_t) \circ d\Xb_t,
\end{equation}
which represent many physical quantities of interest, depending on the choice for the functions $f:\nobreak\RR^d\to \RR$ and $g:\RR^d \to \RR^d$. These include, for example, residence times, the entropy production, and other work-like quantities arising in stochastic thermodynamics~\cite{touchette_introduction_2017, seifert_stochastic_2012}. While the exact probability density $\rho_T(a)$ of $A_T$ cannot be obtained exactly, in general, it is known to scale for large observation times $T$ as
\begin{equation}
\label{eq:ldp}
\rho_T(a) \asymp e^{-T I(a)},
\end{equation}
where the symbol $\asymp$ denotes asymptotic equality up to logarithmic corrections. This result defines the large deviation approximation of $\rho_T(a)$, characterised by the rate function $I:\RR\to\RR$ \cite{dembo1998}. Calculating or estimating this function has become a central problem in statistical physics, as it not only determines the likelihood of fluctuations of $A_T$ around its typical value, but also provides information about the phase behavior and symmetries of nonequilibrium systems~\cite{kohn_magnetic_2005,speck_large_2012, fodor2021irreversibility}.

In most cases, the rate function is obtained not directly from the density of $A_T$, but from the Legendre transform of the scaled cumulant generating function (SCGF) of $A_T$, defined as
\begin{equation}
\label{eq:scgf}
    \psi(\lambda) = \lim_{T\to\infty} \frac1T \log \EE_{\Xb} e^{\lambda T A_T},
\end{equation}
where $\EE_{\Xb}$ denotes an expectation over~\eqref{eq:sde} and $\lambda\in\RR$ is a parameter conjugate to $A_T$. For Markov processes, the SCGF is the dominant eigenvalue of a linear operator, corresponding in the case of diffusions to a modification of the Fokker--Planck generator~\cite{touchette_large_2009}. Hence, the computation of the SCGF and, in turn, the rate function, reduces to a spectral problem, which can be solved if the system's size or dimension is not too large. Alternatively, one can attempt to sample trajectories using path space Monte Carlo to estimate the expectation in the SCGF; however, this approach is not efficient, in general, since it involves exponentially rare events that do not occur spontaneously on timescales accessible to simulations.  

To address these limitations, many strategies have been proposed recently, based on various numerical methods, including the power method~\cite{ferre2018}, diffusion Monte Carlo~\cite{grassberger2002,giardina2006,lecomte2007a,ray2018,ray2020}, recurrent neural network~\cite{Casert_Dynamical_2021}, and reinforcement learning algorithms~\cite{das_variational_2019, rose_reinforcement_2021, das_reinforcement_2021}. The method that we propose is based on importance sampling and proceeds by changing the process $\Xb_t$ to a new process $\Xb_t^u$ governed by the SDE
\begin{equation}
\label{eq:control_sde}
    d\Xbu_t = u_t(\Xbu_t) dt + \sigma d\Wb_t,
\end{equation}
in which the drift $b(\xb)$ is replaced by the control drift $u_t(\xb)$, so as to rewrite the expectation of the SCGF in terms of this new process as
\begin{equation}
\label{eq:tilt}
    \psi(\lambda) = \lim_{T\to \infty} \frac{1}{T}\log \EE_{\Xbu} \left( e^{\lambda T A_T} \frac{d\mathbb{P}[\Xbu]}{d \mathbb{P}_u[\Xbu]} \right).
\end{equation}
The idea with this change of process is to bias the estimation of the expectation towards trajectories that most contribute to the expectation---hence the expression ``importance sampling''---thereby reducing the variance of the simulated estimator. These trajectories are rare with respect to $\Xb_t$; the goal is to make them typical with respect to the new process $\Xb^u_t$. The ratio $d\mathbb{P}[\Xbu]/d \mathbb{P}_u[\Xbu]$ is called the Radon--Nikodym derivative and is there to correct for the fact that the expectation is computed not from the original path probability (or path ensemble) $\mathbb{P}[\Xbu]$, as in \eqref{eq:scgf}, but from a biased path probability $\mathbb{P}_u[\Xbu]$ related to $\Xb^u_t$. This ratio can be computed explicitly along a given path using the Girsanov theorem~\cite{oksendal_stochastic_1992}.

The optimal change of process or optimal control process that achieves the smallest variance in importance sampling is known~\cite{chetrite2015}. Its drift maximizes the cost 
\begin{equation}
\label{eq:lagrangian}
    \mathcal{L}[\Xbu, u] = \lambda TA_T  - \frac12 \int_0^T (u_s-b) D^{-1} (u_s-b) (\Xbu_s) ds,
\end{equation}
which we derive in Appendix~\ref{app:loss}. Moreover, it is known that, in the limit $T\to\infty$, the maximizing control drift is time-independent and that the maximum of the Lagrangian is the SCGF~\cite{chetrite2015}, so that
\begin{equation}
    \label{eq:ritz}
    \psi(\lambda) = \lim_{T\to\infty} \frac1T \sup_{u} \EE_{\Xbu} \mathcal{L}[\Xbu, u].
\end{equation}

This variational representation of the SCGF has a clear interpretation: the first term in the Lagrangian \eqref{eq:lagrangian} enforces the target rare event (constraint) $A_{T} = a$ with a Lagrange multiplier $\lambda$, while the second term is the Girsanov weight related to the change of drift that measures the extent to which the controlled process deviates from the original process \footnote{Alternatively, the optimal control drift can be obtained by contracting rate function characterizing the joint fluctuations of the empirical density and empirical current, commonly known as the ``level-2.5'' large deviation function.}. From this point of view, the optimal control process is interpreted as the process closest to the original process, as measured by the Girsanov weight, that achieves $A_T=a$ as a typical rather than a rare event. In a more physical way, we can also interpret the optimal drift $u^*(\xb)$ of that process as an effective drift that ``creates'' the fluctuation $A_T=a$~\cite{chetrite_nonequilibrium_2014, chetrite2015, jack_effective_2015}. This provide as physical mechanism explaining how fluctuations are created in time, which is useful for studying dynamical phase transitions.

\section{Algorithm}

\begin{algorithm*}[t!]
\caption{Concurrent Training \label{alg:concurrent}} \begin{algorithmic}[1]
    \State{{\bfseries Data:} Lagrangian $\mathcal{L}[\Xbu, \delta u(\Xbu_t,\lambda; \thetab), \lambda]$, initial $\thetab$, $k_\text{max}\in\NN$ total duration, $T\in\RR$ the duration of sampled trajectory, $N(m)\in\NN$ the batch size. $\{\lambda_1, \lambda_2,\cdots,\lambda_M\}\subset\RR$, $\alpha > 0$ the learning rate.}
    \State{$k=0$}
    \While{$k<k_\text{max}$}
    \For{$m=1,\dots, M$}
    \For{$n=1,\dots, N(m)$}
    \State{Sample $\Xb_{[0,T], n}^{u(\lambda_m)}$ according to $d\Xbu_t = [b(\Xbu_t) + \delta u(\Xbu_t, \lambda_m; \thetab)]dt + \sigma d\Wb_t$ with initial condition $\Xb_{0, n}^{u(\lambda_m)}$};
    \EndFor
    \EndFor
    \State{Compute
    \begin{align}
        &\mathcal{L}_{\thetab}^{(m,n)} =  \lambda_n\int_0^T f(\Xb_{t, n}^{u(\lambda_m)})dt + g(\Xb_{t, n}^{u(\lambda_m)})\circ d\Xb_{t, m}^{u(\lambda_m)} - \frac{1}{2}\int_0^T \delta u(\Xb_{t, n}^{u(\lambda_m)},\lambda_m;\thetab) D^{-1} \delta u(\Xb_{t, n}^{u(\lambda_m)},\lambda_m;\thetab) dt \nonumber \\
        &\nabla_{\thetab}\mathcal{L}(\thetab) = \frac{1}{M}\sum_{m=1}^M\frac{1}{N(m)}\sum_{n=1}^{N(m)}\nabla_{\thetab}\mathcal{L}_{\thetab}^{(m,n)} \nonumber
    \end{align}}
    \State{Update $\thetab \gets \thetab + \alpha\nabla_{\thetab}\mathcal{L}(\thetab)$}
    \State{Update the initial condition $\Xb_{0, n}^{u(\lambda_m)}\gets \Xb_{T, n}^{u(\lambda_m)}$} 
    \Procedure{(Optional) Replica exchange:}{} 
        \State Select two random integers $n_1, n_2$ such that $n_i\leqslant N(m_i)$;
        \State Compute the Radon-Nikodym derivative:
    \begin{align}
        M_T = \frac{d\PP[\Xb_{[0,T],n_1}^{u(\lambda_{m_1})}]}{d\PP[\Xb_{[0,T],n_2}^{u(\lambda_{m_2})}]} = \frac{\exp\left\{-\frac{1}{4D}\int_0^T |\dot{\Xb}_{t,n_1}^{u(\lambda_{m_1})} - u(\Xb_{t,n_1}^{u(\lambda_{m_1})}, \lambda_{m_1}) |^2dt\right\}}{\exp\left\{-\frac{1}{4D}\int_0^T |\dot{\Xb}_{t,n_2}^{u(\lambda_{m_2})} - u(\Xb_{t,n_2}^{u(\lambda_{m_2})}, \lambda_{m_2}) |^2dt\right\}} \nonumber
    \end{align}
    \State{$u\sim\mathrm{Uniform}(0,1)$}
    \If{$u<\min[1,M_T]$}
        \State{exchange $\Xb_{0, n_1}^{u(\lambda_{m_1})}$ and $\Xb_{0, {n_2}}^{u(\lambda_{m_2})}$.}
    \EndIf
    \EndProcedure
    \EndWhile
\State{{\bfseries return:} $\thetab$}.
\end{algorithmic}
\end{algorithm*}

The variational representation of the SCGF shown in \eqref{eq:ritz} has a form that is standard in control theory and, as such, is amenable to Ritz-type methods that optimize a parametric representation $u(\xb, \lambda; \thetab)$ with respect to some set of variational parameters $\thetab$. Directly carrying out this optimization is nontrivial, as it requires representing a potentially complex, many-body force, motivating several sophisticated strategies that rely on intricate basis functions, Malliavin weight sampling, and reinforcement learning~\cite{das_variational_2019, rose_reinforcement_2021, das_reinforcement_2021, oakes2020}.

Our contribution is to solve this high-dimensional control problem using gradient-based optimization and deep neural networks, which are well-suited to this task~\cite{rotskoff_parameters_2018, chizat_global_2018, mei_mean_2018, sirignano_mean_2018, barron_universal_1993, cybenko_approximation_1989} due to their robust function approximation properties, even in high-dimensional settings.  The pseudo-code of our optimization algorithm is presented in Algorithm~\ref{alg:concurrent} and a Python source code is available online \footnote{The Python source code is available online at \href{https://github.com/quark-strange/machine_learning_LDP}{github.com/quark-strange/machine\_learning\_LDP}}. There are four important components to our algorithm:

\paragraph{Neural network representation of the drift.} Following recent works on the deep Ritz method~\cite{e_deep_2018}, we represent the change in control drift
\begin{equation}
\delta u(\xb) = u(\xb)-b(\xb) 
\end{equation}
using a neural network that contains multiple layers $L_i$, where each layer consists of two linear transformation, two nonlinear activation functions and a residual connection:
\begin{equation}
    L_i(\Xb) = \phi[W_{i,2}\cdot\phi(W_{i,1}\Xb+b_{i,1})+b_{i,2}]+\Xb,
\end{equation}
where $W_{i,j}\in\RR^{h\times h}$ and $b_{i,j}\in\RR^{h}$ are parameters for the $i$-th layer, $h$ is the dimension of the hidden layers, and $\phi$ is the activation function. The residual connection expressing each layer as $L_i(\Xb) = f(\Xb) + \Xb$ helps with stability and avoiding the vanishing gradient problem. 

Since our approach requires simulating trajectories from \eqref{eq:control_sde}, an unbounded activation such as ReLU may lead to divergence of the sampled trajectories. To avoid this problem, we use $\tanh(\cdot)$ as the activation function throughout this paper though other nonlinearities may also be suitable. 
The full network can then be expressed as 
\begin{equation}
    z_{\thetab}(\Xb) = L_n \otimes \dots \otimes L_1(\Xb).
\end{equation}
The input $\Xb\in\RR^d$ for the first layer is padded by a zero vector when $d<h$. Finally, the ansatz $\delta u(\Xb_t, \lambda; \thetab)\in\RR^d$ is expressed as a linear transform of $z_{\thetab}(\Xb)$. 

\begin{figure*}
    \begin{center}
    \includegraphics[width=.95\linewidth]{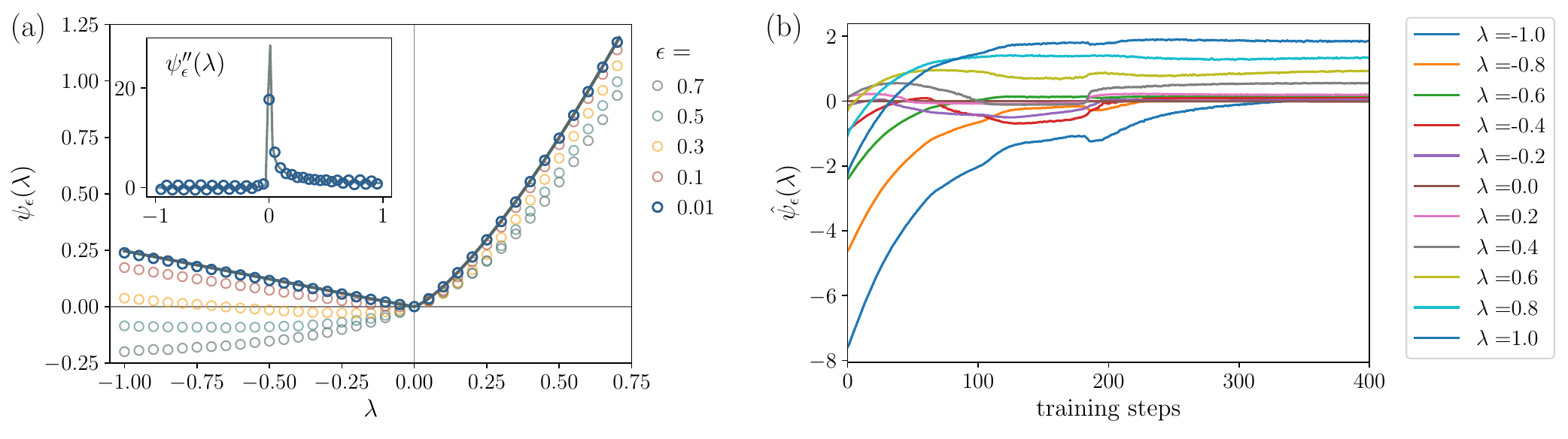}
    \end{center}
    \vspace{-7mm}
    \caption{Results of the diffusion in the quartic potential. (a) SCGF for the observable \eqref{eq:quarticobs} for decreasing temperatures $\epsilon$. The solid line represents the exact solution \eqref{eq:quarticexact} in the zero-noise limit. The inserted figure shows the second derivative of the SCGF for $\epsilon=0.01$, confirming a second order dynamical phase transition. In this example the hidden layer dimension and number of layers of the neural network are 50 and 2, respectively. A smaller hidden layer dimension such as 10 is able to generate results with similar accuracy but requires longer time for training. We first select 11 $\lambda$ values uniformly from $-1$ to $1$, where each $\lambda$ contains 20 replica. At each training step, a total number of 220 trajectories with $T=5$ are generated by the Euler-Maruyama method ($dt=10^{-3}$). The neural network is updated through standard back propagation where the gradient is computed by the adaptive gradient algorithm method (AdaGrad) with a learning rate $5\times10^{-3}$. The resulting estimation of the SCGF is then refined by changing $\lambda$ and by simulating the resulting driven process. (b) Illustration of the concurrent training with $\epsilon=0.01$. Each line corresponds to the evolution of the cost function with a specific $\lambda$.}
    \label{fig:LNL}
\end{figure*}

\paragraph{Loss estimation and gradient.} The loss function is estimated, for a given change of drift $\delta u(\cdot,\lambda;\thetab)$, with a collection or ``batch'' of $N$ trajectories generated in parallel using direct Langevin dynamics. The variance and convergence of the resulting estimator are discussed in Appendix~\ref{app:var}, which shows that short time trajectories suffice when the batch size is large. 

From the estimated loss, we proceed to compute the loss gradient to update the parameters $\thetab$ by differentiating through the solution of the SDE~\eqref{eq:control_sde} using recent developments in machine learning~\cite{tzen_neural_2019, li_scalable_2020}. Over short times, we use direct back-propagation of the dynamics through a Stratonovich time-discretization of the SDE to compute $\grad_{\thetab}\mathcal{L}$. The computational graph that contains all the gradient information consumes significant memory resources in this case, so over longer timescales, we calculate $\grad_{\thetab}{\mathcal{L}}$ by solving instead an adjoint SDE, detailed in Appendix~\ref{app:adjoint}. This method is stable and only requires that we keep the noise history and solve the SDE backward in time. 

\paragraph{Estimation of the SCGF and rate function.} The repeated gradient minimization of the loss yields, after enough gradient steps, a single estimated point $\psi(\lambda)$. To obtain the rate function, the SCGF must be estimated by training the neural network for multiple values of $\lambda$ either simultaneously or sequentially. In the first case, which we term concurrent training, the loss function at each training step is evaluated as the mean of the loss function with each $\lambda_m$ from a set $\{\lambda_1, \lambda_2,\cdots,\lambda_M\}$. We find that the expressiveness of the neural networks we use allows a single force function $u(\cdot, \lambda)$ to capture the control forces for a wide range of $\lambda$, even when there are multiple dynamical phases. For high-dimensional systems, where the batch size is limited, one may alternatively start with a given $\lambda$, e.g., 0, and sequentially increase or decrease $\lambda$. This sequential training approach, which is similar to transfer learning~\cite{li_fourier_2021}, shows fast convergence. 

\paragraph{Replica exchange.} Near dynamical phase transitions, which lead to rapid changes of the optimal control forces as a function of $\lambda$, we have found that it is useful to share information from distinct values of $\lambda$ by employing a path space variant of the replica exchange method~\cite{frenkel_understanding_2002}, in which two trajectories corresponding to different $\lambda$ are swapped according to a Metropolis-Hastings algorithm that uses the loss function in place of an energy. This increases the likelihood of sampling trajectories in different phases, leading to a more accurate estimation of the SCGF.

\section{Applications}

\begin{figure*}
    \centering
    \includegraphics[width=.95\textwidth]{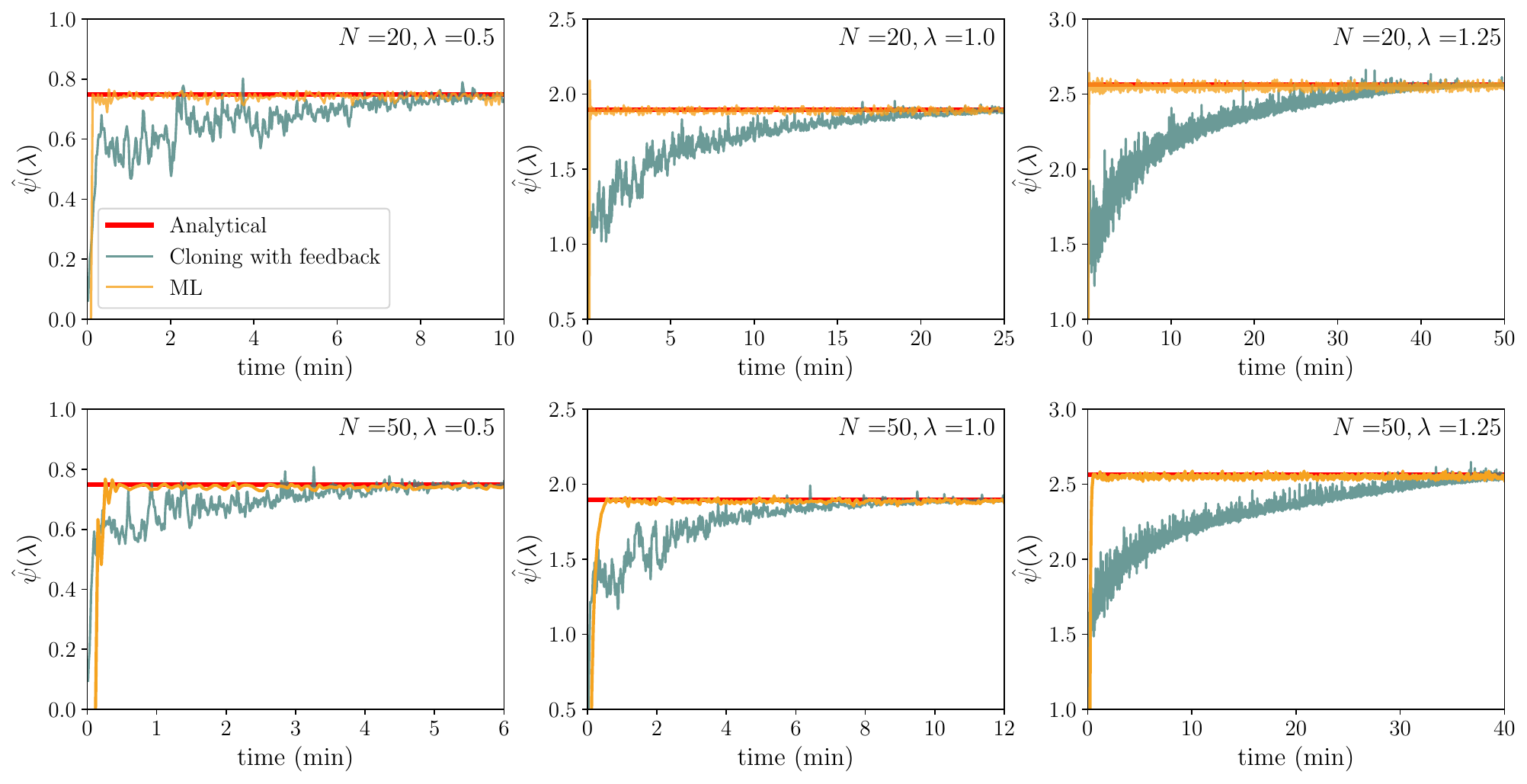}
    \caption{Comparison of our machine learning approach with the cloning method with feedback. We computed the SCGF for the 1D diffusion system described as in \eqref{eq:1d_diffusion} with $\epsilon=0.01$. $N$ is the batch size (for the ML approach) or the number of clones (for the cloning method). For the cloning method, we set the trajectory length to be $T=0.3$ ($dt=10^{-3}$), and updated the control potential every 75 steps so the time interval between updating is 22.5. All results were computed in the same personal computer, using CPU only.}
    \label{fig:comparison}
\end{figure*}

We test our algorithm  on two models which have been studied before in the context of large deviations and which illustrate two different challenges faced by large deviation numerical methods, namely: critical slowing-down effects related to dynamical phase transitions, and the representation of the control force for high-dimensional systems, in particular, many-body systems.

\subsection{Simple diffusion}

For the first test, we consider a 1D diffusion in a quartic potential,
\begin{equation}
    \label{eq:1d_diffusion}
    dX_t = -X_t^3 dt + \sqrt{2\epsilon} dW_t,
\end{equation}
and focus on the observable
\begin{equation}
   \label{eq:quarticobs}
    A_T = \frac{1}{T}\int_0^T X_t (X_t + 1)dt.
\end{equation}
For this model, the SCGF scaled by the strength $\epsilon$ of the noise is known to display a second-order dynamical phase transition in the small-noise limit, meaning that the derivative of $\psi_\epsilon(\lambda)=\epsilon \psi(\lambda)$ is not differentiable at $\lambda=0$ when considering the additional limit $\epsilon\to 0$. This can be checked from the exact result
\begin{equation}
   \label{eq:quarticexact}
   \psi_0(\lambda) = \lim_{\epsilon\to 0} \psi_\epsilon(\lambda) = \max_{q}\{\lambda(q^2+q)-q^6/4\}.
\end{equation}
Resolving this phase transition using cloning algorithms is challenging, due to a critical slowing down of the dynamics, which can be alleviated to some degree by incorporating adaptive feedback methods~\cite{nemoto_populationdynamics_2016}. 

The low-noise limit is not a bottleneck for our algorithm. Using short trajectories ($T=5$), we concurrently trained a single neural network with a set of values for $\lambda$ in the range $[-1,1]$. The results, plotted in Fig.~\ref{fig:LNL}(a), agree exceptionally well for $\epsilon=0.01$ with the exact result obtained in the low-noise limit. For most values of $\lambda$, we find in fact that the normalized mean squared error between our estimate of the SCGF and the exact result is about $0.2\%$. This can be reduced by training the network for a single $\lambda$ rather than concurrently for many $\lambda$ values. Replica exchange is not crucial here and does not noticeably improve the accuracy. The numbers of steps required to reach $\psi(\lambda)$ is shown in Fig.~\ref{fig:LNL}(b) to vary little for different $\epsilon$---typically in the range of 400 to 600 steps. The rapid convergence that we observe away from the dynamical phase transition may be due to the fact that we employ overparameterized neural networks, which do not suffer from overfitting and converge to global minimizers when the loss function can be repeatedly sampled, a setting known as online learning~\cite{rotskoff_trainability_2018, chizat_global_2018,belkin_reconciling_2019}.

To compare our algorithm with the cloning algorithm with feedback, we have applied the latter to the same model. In brief, the cloning method~\cite{giardina2006} evaluates the SCGF by simulating a batch of trajectories (clones) and by duplicating or eliminating trajectories according to weights computed from the trajectory ensemble.
Generally, this population dynamics method requires a exponentially large number of replicas of the system as the desired event becomes rarer (or equivalently, as the magnitude of the noise decreases). To overcome this issue, \cite{nemoto_populationdynamics_2016} proposed a feedback approach, in which a controlling potential function is adaptively constructed to modify the original dynamics. 

The convergence of this cloning algorithm with feedback with that of our algorithm for various batch or clone sizes and values of $\lambda$ are compared in Fig.~\ref{fig:comparison} in terms of computational time on a single machine measured in minutes. The results clearly show that our algorithm is significantly faster and more stable than the cloning algorithm, especially when the biasing parameter $\lambda$ is far from 0. The difference in performance is partly due to the fact that the feedback in the cloning algorithm relies on estimating two probability distributions, which is limited by the relaxation time of the original system. Moreover, as $\lambda$ deviates from 0, more iterations for updating the control potential are required to realize the rare events. In our algorithm, no distributions are estimated: the control drift is obtained directly by the taking the gradient of the estimated cost over a number of batches, which, contrary to the cloning algorithm, need not be stored in memory. Moreover, the results plotted in Fig.~\ref{fig:comparison} show again that the number of steps needed to converge to the optimal drift does not vary much with $\lambda$.

There is another significant difference in that the control potential in the cloning algorithm is represented by a linear combination of a set of basis functions such as polynomials, so it requires \textit{a priori} knowledge in order to choose the adequate basis functions, often in a case by case manner. By comparison, the machine learning approach that we propose is agnostic, meaning that it can be applied to a broad class of problems without any modifications of the algorithm or specific knowledge of the underlying structure of the problem \footnote{While it is not the case for this simple example, it may be possible for the cloning method to outperform the machine learning approach if the basis is expertly tailored to the problem at hand.}. Yet another advantage is that it has an inherently parallel structure and can evaluate the SCGF for multiple $\lambda$ simultaneously. Cloning does not benefit from this parallel structure as the SCGF must be evaluated with a potential specific to each value of $\lambda.$

\subsection{Active Brownian particles}

\begin{figure}[t]
    \centering
    \includegraphics[width=.95\linewidth]{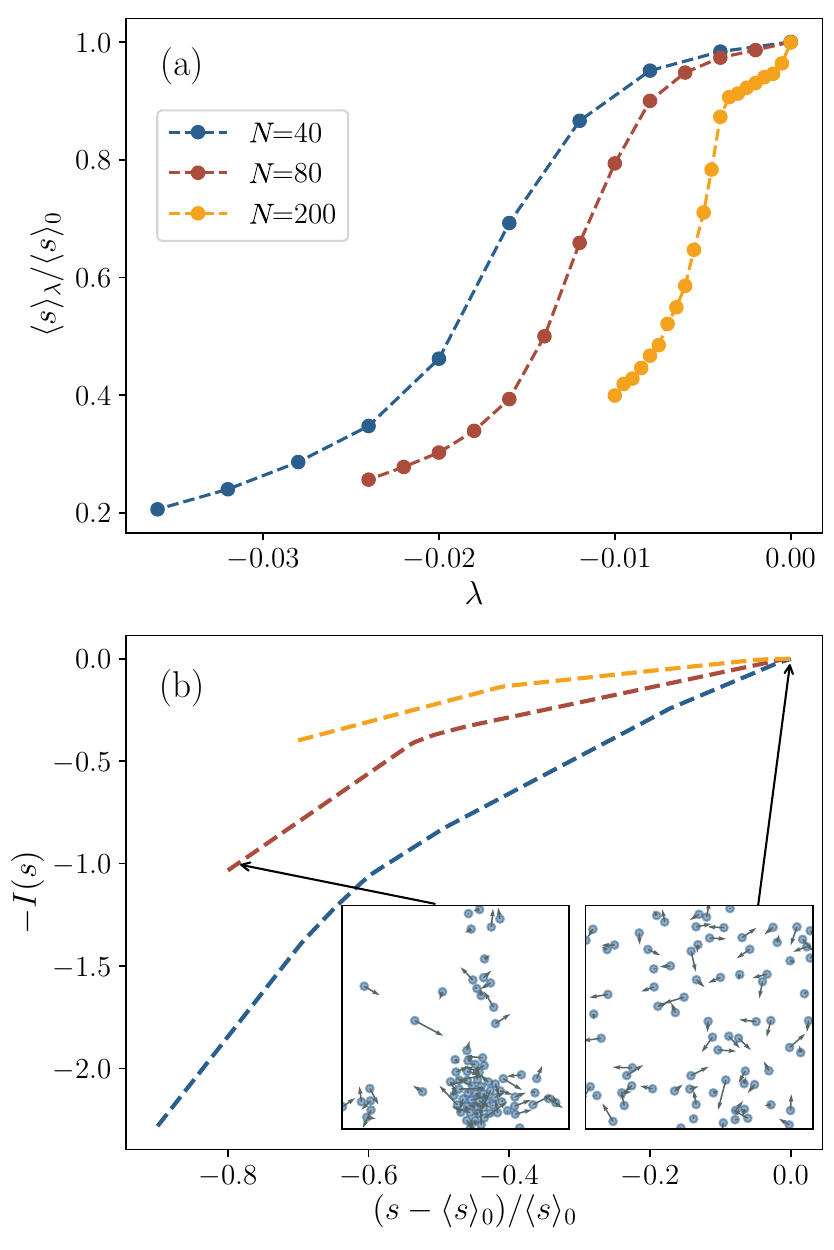}
    \caption{
    Small entropy production indicates particle clustering. (a): The average entropy production at given $\lambda$ for different system sizes: $N=40$ (blue lines), 80 (red lines), and 200 (yellow lines). (b): The corresponding rate functions. The inserted figures show snap shots of typical behaviors in the high entropy production phase ($\lambda=0$) and low entropy production phase ($\lambda=-0.05$), respectively. The arrow represents the direction of motion. In this example the hidden layer dimension and number of layers of the neural network are 1000 and 6 respectively. The batch sizes for 40 and 80 particles are 75, and 20 for the 200 particle case. $T=0.1$ and $dt=10^{-4}$. The density of particles throughout all three cases is $\rho = N/L^2=0.1$ where $L$ is the length of the simulation box. To avoid boundary effect, the input of the neural network is not the absolute position of particles but its relative position to a particular one, i.e., $u(\{\Xb^{(i)}-\Xb^{(0)}\})$ instead of $u(\{\Xb^{i}\})$. This step is essential otherwise the learned control force would force particles to the boundary.}
    \label{fig:abp}
\end{figure}

Theoretical~\cite{cates_motilityinduced_2015,bialke_effective_2014,pietzonka_extreme_2016} and numerical~\cite{hagan_structure_2013} characterizations of active matter provide a compelling model for nonequilibrium phenomena. Minimal models, such as actively driven Brownian particles (ABPs) with purely repulsive interaction potentials, exhibit a rich spectrum of collective fluctuations and nonequilibrium phase separation emerging from the impact of persistent, directional motion on the local diffusivity of the constituent particles. 
The precise connection between energy dissipation and pattern formation in these nonequilibrium transitions remains a topic of intense research~\cite{nguyen_design_2016, fodor_how_2019, fodor_dissipation_2020}. For example, the correlation between the structure formation in ABPs and fluctuations in entropy production was recently described by \citet{grandpre_entropy_2021}.
Probing the connection between rare dynamical behavior and collective fluctuations, however, is extremely challenging because the onset of clustering in ABPs requires large system sizes and high densities that can be accessed by cloning type algorithms only with a large number of replicas.

To test our algorithm, we consider the ABP model in which the motion of the $i$th particle is governed by the following equations:
\begin{equation}
\begin{aligned}
    \label{eq:abp}
     &d\Xb_t^{(i)} = [- \mu\diff{U(\Xb_t)}{\xb^{(i)}} + v\bb^{(i)}_t]dt + \sqrt{2D}d\Wb_t^{(i)}, \\
     &\bb^{(i)} = [\cos\phi_t^{(i)}, \sin\phi_t^{(i)}]^\top,\ d\phi_t^{(i)} = \sqrt{6D}dW_t^{\phi^{(i)}}.
\end{aligned}
\end{equation}
The potential $U(\Xb_t)$ defining the conservative interparticle force is taken here to be a purely repulsive Weeks-Chandler-Andersen (WCA) pair potential that depends on the relative distance $l_{ij}$ according to:
\begin{equation}
    U(l_{ij}) = \left\{\begin{array}{lcl}
        4\epsilon\left[ (\sigma/l_{ij})^{12} - (\sigma/l_{ij})^{6}  \right] + \epsilon, & l_{ij}\leqslant 2^{1/6}\sigma \\
        0, & l_{ij} > 2^{1/6}\sigma.
    \end{array} \right. 
\end{equation}
The non-conservative self-propulsion term $v\bb^{(i)}$ represents the dissipative ``active'' force in which $\bb_t^{(i)}$ are unit vectors that rotate diffusively and $v$ is the magnitude of the active force. Finally, $\Wb_t^{(i)}$ and $W_t^{\phi^{(i)}}$ are independent standard Wiener processes representing noise sources for the state and angle. The simulations are performed with periodic boundary condition, and the relative distance matrix $l_{ij}$ is adjusted by the minimum image convention. The unit of length is also normalized by $\sigma$ and we set $\epsilon=1$.

The phase separation properties of this model have been studied extensively~\cite{cates_motilityinduced_2015}. When the P\'{e}clet number and the density of particles are high enough, the system exhibits a motility induced phase transition in which a macroscopic aggregate of particles forms. This transition has a natural dynamical correlate with the average entropy production
\begin{equation}
    \label{eq:ep}
    s = \frac{1}{NT}\sum_{i=1}^N\int_0^T v\bb^{(i)}_t D^{-1} \circ d\Xb_t^{(i)}.
\end{equation}
When the system enters the phase separated state, much of the directional motion also ceases, leading to a drop in the average entropy production compared to an unclustered trajectory. Indeed, several studies have pointed to entropy production being a natural observable for studying motility-induced phase separation~\cite{grandpre_entropy_2021} and nonequilibrium pattern formation in liquids~\cite{fodor_dissipation_2020,fodor_how_2019}, though a control-based approach has not been pursued for these systems to date. 

Using our algorithm, we computed the many-body control forces associated with fluctuations of the entropy production for various particle numbers ($N=40, 80, 200$). The results, shown in Fig.~\ref{fig:abp}, were computed through the sequential training, since concurrent training requires a large total batch size which is computationally costly for high dimensional systems. For this system, it is crucial that we do not include the direction of the active particles $\phi_t$ in the state, otherwise the entropy production rate can trivially be reduced by learning control forces anti-parallel to the direction of the active force; this choice has a physical justification, namely, the directions are in equilibrium and are not reversed under time-reversal. 

\begin{figure}
    \centering
    \includegraphics[width=0.95\linewidth]{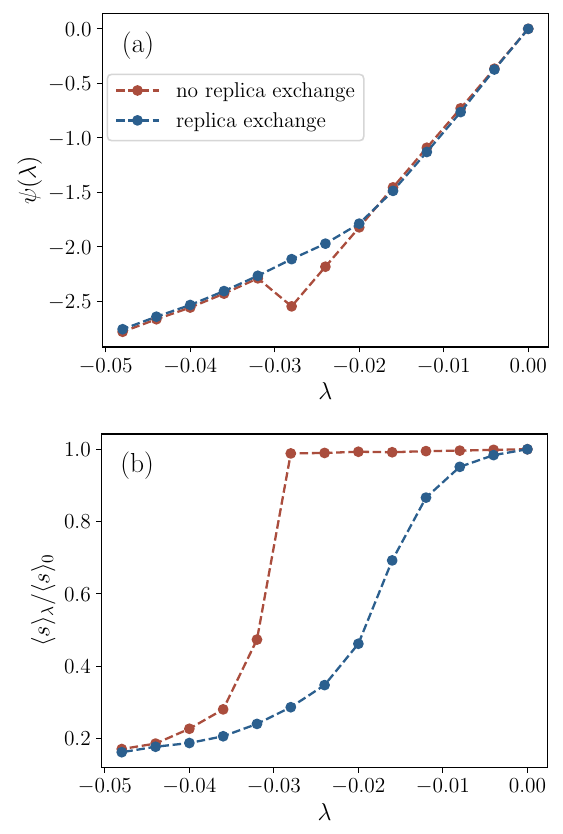}
    \caption{Estimate of the SCGF and corresponding changes of average entropy production for $N=40$ with and without replica exchange. The blue line corresponds to the results with replica exchange and the red line shows the results without.}
    \label{fig:replica_exchange}
\end{figure}

The simulations converge over relatively long times when first driving the system into the clustering phase; however, once we obtain a control force, they converge fast when sequentially altering $\lambda$.  
For $N=40,80$, we also noticed that replica exchange is required to obtain a convex SCGF, whereas for $N=200$, replica exchange is not necessary. The replica exchange is implemented by concurrently training with multiple $\lambda$ values and by swapping trajectories between them. We set $\lambda_0=-0.05$ with batch size 75. Then at each step all the 75 trajectories are attempted to be exchanged, as explained in Algorithm~\ref{alg:concurrent}. In Fig.~\ref{fig:replica_exchange} we plot the results with and without replica exchange, respectively, in the $N=40$ case. The results indicate that replica exchange is essential for obtaining a convex SCGF.

Going back to our results in Fig.~\ref{fig:abp} (see also the supplementary movie), we can see that particles start to aggregate when the biasing field $\lambda$ is sufficiently negative. For all system sizes, the entropy production rate changes dramatically as a function of $\lambda$ around a value coinciding with the onset of clustering. This sharp transition signifies a first-order dynamical phase transition in the entropy production rate, shown in Fig.~\ref{fig:abp}(b), related to a singularity in the rate function at the transition point. Examining the learned controls provides further insight into the microscopic origins of the transition. As shown in the inset of Fig.~\ref{fig:abp}(b), the learned control forces lead to net forces on the particles that favor the aggregated state.

The nonequilibrium fluctuations of active systems have been studied in a variety of contexts, using unbiased sampling~\cite{caprini_fluctuation_2021}, cloning~\cite{grandpre_entropy_2021}, and reinforcement learning~\cite{das_variational_2019}. Our approach considerably simplifies the computation compared to reinforcement learning because we do not need to learn an expected value function.
Moreover, unlike cloning, our approach scales to high-dimensional systems without incurring significant additional computational cost; training for various $\lambda$ is easily parallelizable and the integration of the trajectories can be carried out on heterogeneous hardware. Indeed, the cloning algorithm would require prohibitive computational resources compared with our algorithm.

\section{Discussion}

The results presented in this paper demonstrate the efficacy of a machine learning algorithm that adaptively learns optimal control forces to directly estimate large deviation functions for systems extremely challenging for conventional methods. The algorithm that we have proposed relies on direct stochastic optimization based on a small number of trajectories, which themselves may not need to have a long duration--a fact that requires further investigation. Importantly, the Lagrangian that we optimize is explicit and exact in the long time limit, requiring no additional approximation or optimization, as only the control function is learned. We have shown that the approach is robust both near the dynamical phase transitions and in the limit of small noise.
Like many methods based on machine learning, the method we propose shows favorable performance in high-dimensional systems and still identifies many-body control forces that realize rare fluctuations near dynamical phase transitions.   

The examples we have explored are continuous-time stochastic differential equations with a constant diffusion term (and hence additive noise), but it is straightforward to adapt our algorithm to other types of systems, including those with multiplicative noise, or with discrete, but innumerable state spaces such as unbounded Markov jump processes where directly evaluating the principal eigenvalue is not possible. 
Our approach could also be extended to finite-time large deviations, though we anticipate that this would require longer trajectories and therefore the adjoint state method would likely be mandatory. 
Learning control forces that drive the system locally, and hence can be transferred to systems of increasing size and complexity is among the most attractive possibilities for future investigation.
For interacting particle systems, if the form of the input and the architecture of the neural network are carefully designed, it may be possible to obtain the optimal control force for systems with thousands of particles by training on smaller, more computationally tractable systems.

\begin{acknowledgments}
G.M.R.\ acknowledges support from the Terman Faculty Fellowship. The authors also thank Eric Vanden-Eijnden and Suri Vaikuntanathan for thoughtful comments on this manuscript.
\end{acknowledgments}

\appendix 

% \onecolumngrid

\section{Derivation of the cost functional}
\label{app:loss}

Using importance sampling, we write the expression for the SCGF as an expectation over a ``tilted'' or biased path measure,
\begin{equation}
\label{eq:is_psi}
 \psi(\lambda) = \lim_{T\to\infty} \frac{1}{T} \log \int e^{\lambda T A_T[\Xbu]} \frac{d\mathbb{P}[\Xbu]}{d\mathbb{P}_u[\Xbu]} d\mathbb{P}_u[\Xbu].
\end{equation}
This expectation must be estimated for each $\lambda$ of interest by collecting trajectories from the controlled process~(6).
The relative statistical weight of the unperturbed path measure $\mathbb{P}$ to the path measure of the controlled process $\mathbb{P}_u$ is defined through the Radon--Nikodym derivative.
In our case, using the parameterization $u(\xb,\lambda) = b(\xb) + \delta u(\xb,\lambda)$, this derivative can be written explicitly using the Girsanov theorem~\cite{oksendal_stochastic_1992} as
\begin{widetext}
\begin{equation}
    \frac{d\mathbb{P}[\Xbu]}{d\mathbb{P}_u[\Xbu]} \equiv M_T = \exp \left( -\int_0^T \sigma^{-1} \delta u(\Xbu) dW_t - \frac12 \int_0^T \delta u(\Xbu_t) D^{-1} \delta u(\Xbu_t) dt \right),
\end{equation}
where we use the notation $M_T$ to emphasize the fact that $M_T$ is a martingale. The first integral in the exponential can be neglected when the deterministic contribution is finite and we are left with an expression for~\eqref{eq:is_psi}:
\begin{equation}
 \psi(\lambda) = \lim_{T\to\infty} \frac{1}{T} \log \EE_{\Xbu} \exp \left(\lambda T A_T[\Xbu]  - \frac12 \int_0^T \delta u(\Xbu_t) D^{-1} \delta u(\Xbu_t) dt \right).
\end{equation}

The term inside the exponential is evidently time-extensive and, in the limit $T\to \infty$, the integral will be dominated by a saddle point, following the Laplace approximation. 
As a result, we obtain
\begin{equation}
 \psi(\lambda) = \lim_{T\to\infty} \frac{1}{T} \sup_{\delta u} \EE_{\Xbu}  \left\{ \lambda T A_T[\Xbu]  - \frac12 \int_0^T \delta u(\Xbu_t) D^{-1} \delta u(\Xbu_t) dt \right\}.
\end{equation}
Hence, the argument of the supremum becomes a natural variational objective for $\delta u$, which we denote 
\begin{equation}
    \mathcal{L}[\Xbu, u] = \lambda \int_0^T f(\Xbu_t) dt + g(\Xbu_t)\circ d\Xbu_t - \frac12 \int_0^T \delta u D^{-1} \delta u (\Xbu_t) dt.
\end{equation}
\end{widetext}

\section{Cost estimator}
\label{app:var}

\begin{figure*}[t]
    \centering
    \includegraphics[width=.9\linewidth]{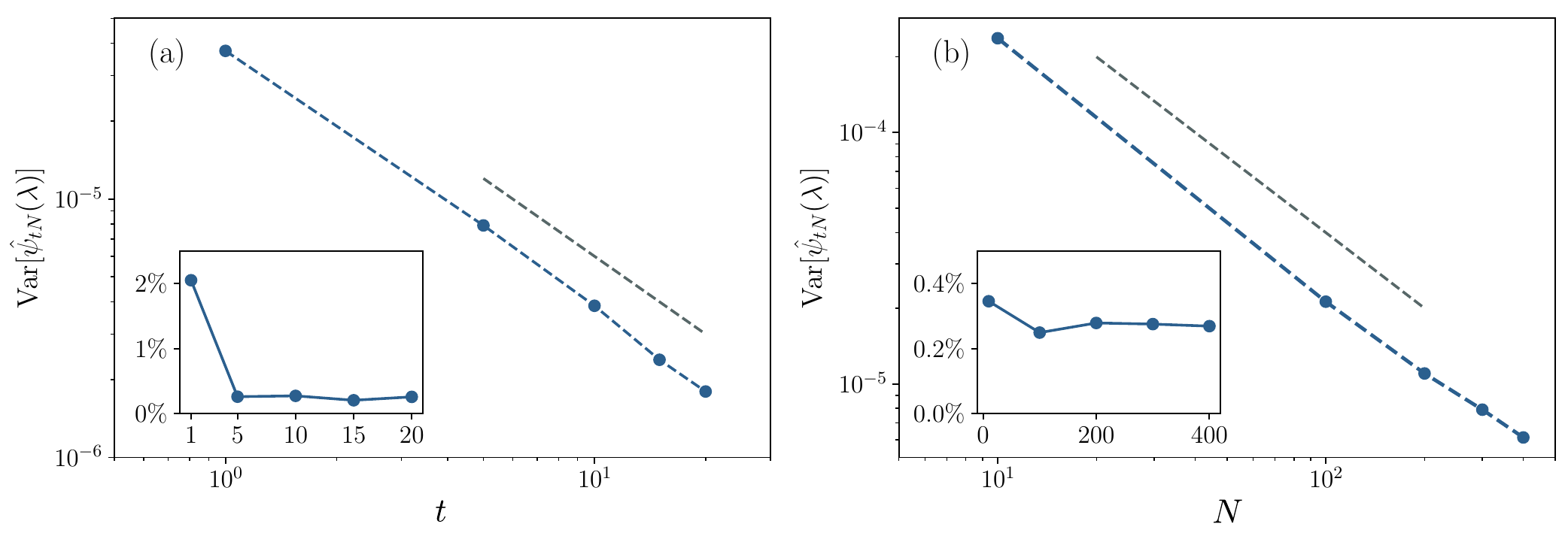}
    \caption{Variance of the estimator. We illustrate the scaling property of the variance of our estimator \eqref{eq:estimator} using the small noise example. By fixing $\lambda=1$ and $\epsilon=0.01$, the neural network is trained with (a) a fixed batch size $N=300$ with various trajectory length t from 1 to 20, or (b) a fixed $t=5$ and various batch size from 10 to 400. The neural networks in all cases are trained for more than 400 steps, and the variance is estimated by collecting the data from the last 100 steps. The insert figures show the relative absolute error $|\hat{\psi}(\lambda) - \psi(\lambda)|/\psi(\lambda)$. The gray dashed line represents a $-1$ slope.}
    \label{fig:LNL_variance}
\end{figure*}

We compute the cost functional numerically by simulating $N$ independent trajectories $\Xbu_{t,i}$ of the controlled process, referred to as replica, over a finite time-window or horizon $[0,t]$ by using the estimator
\begin{equation}
    \label{eq:estimator}
    \hat{\psi}_{Nt}(\lambda) =\frac{1}{Nt} \sum_{i=1}^N \hat{\psi}_{t,i}(\lambda),
\end{equation}
where
\begin{widetext}
\begin{equation}
    \hat{\psi}_{t,i}(\lambda) = \frac{1}{t}\left\{\lambda \int_0^t f(\Xbu_{s,i}) ds + g(\Xbu_{s,i})\circ d\Xbu_{s,i}  - \frac12 \int_0^t \delta u(\Xbu_{s,i}) D^{-1} \delta u(\Xbu_{s,i}) ds\right\}
\end{equation}
\end{widetext}
is the estimator of the cost functional for one replica. By the ergodic theorem and the law of large numbers, $\hat{\psi}_{Nt}(\lambda)$ converges to the SCGF $\psi(\lambda)$ in the double limit $t\to\infty$ and $N\to\infty$, provided that $u$ is the optimal control drift $u^*$.  Note however that, since $u^*$ is time-independent, we can obtain the long-time limit of the optimal cost by considering a finite-time estimator provided that we take the limit $N\to\infty$, so that there is only one limit to consider.

This point is illustrated for the 1D diffusion in the low-noise limit in Fig.~\ref{fig:LNL_variance}, which shows the mean square error (MSE) of the loss estimator or, equivalently, its variance since it is unbiased:
\begin{equation}
\mathrm{MSE}=\EE[\hat{\psi}_{Nt}(\lambda) - \psi(\lambda)]^2 = \Var[\hat{\psi}_{Nt}(\lambda)].
\end{equation} 
Since the $N$ replica are independent, the variance of $\hat{\psi}_{Nt,i}(\lambda)$ must scale with $N^{-1}$ due to the central limit theorem, yielding $\mathrm{MSE} = \Var[\hat{\psi}_{t,i}(\lambda)]/N$.
In general, $\hat{\psi}_{t,i}(\lambda)$ itself is a time-extensive variable that satisfies a large deviation principle, so its variance $ \Var[\hat{\psi}_{t,i}(\lambda)]$ scales with $t^{-1}$. Therefore, overall, the MSE of the loss estimator decreases as $(tN)^{-1}$. Hence, a large-batch and short-time estimator is equivalent to a small-batch and long-time estimator. In Fig.~\ref{fig:LNL_variance}, we confirm this scaling property by plotting the estimator of $\psi_\epsilon(\lambda)$ for the simple diffusion model as a function of integration time or batch size.

\section{Adjoint state method}
\label{app:adjoint}

The adjoint state method for Stratonovich SDEs \footnote{The choice of Ito or Stratonovich convention is immaterial in our examples because we consider only SDEs with additive noise.} differs only marginally from the classical adjoint method for ODEs, though we note that the method can be extended to multiplicative noise~\cite{li_scalable_2020}. These methods require forward and backward integration of the differential equation and, in the stochastic case, one must solve the SDE backward in time with the same Wiener process $\Wb_t$ used in the forward direction, meaning that the noise history must be stored. We explain the method for the ODE case and refer to Ref.~\cite{li_scalable_2020} for further details. 

Consider the ODE
\begin{equation}
 \Diff{x}{t}{} = u(x, t, \theta); \qquad x(0) = x_0
 \label{eq:ode}
\end{equation}
and some objective function $\mathcal{L}(x(T))$, which we would like to minimize with respect to $\theta.$ 
We note that $\mathcal{L}$ depends on $\theta$ through the dynamics because
\begin{equation}
    x(T) = x_0 + \int_0^T u(x,t,\theta) dt.
\end{equation}
The dependence of $\mathcal{L}$ on $\theta$ can be computed using classical sensitivity analysis techniques.
Assuming that we can easily evaluate the cost functional at the final integration time $T$, we need to compute 
\begin{equation}
    \diff{\mathcal{L}(x(T))}{\theta}{} =  \diff{\mathcal{L}(x(T))}{x(T)}{}  \diff{x(T)}{\theta}{},
\end{equation}
where $x(t)$ is constrained to follow the dynamics~\eqref{eq:ode}.
Using the method of Lagrange multipliers, we can turn this into an unconstrained optimization where the time-dependent multiplier $\mathfrak{A}(t)$ is chosen to impose the constraint $\dot{x}=u$.
That is, the cost functional becomes
\begin{equation}
    \tilde{\mathcal{L}}(x(T)) = \mathcal{L}(x(T)) - \int_0^T \mathfrak{A}(t) (\dot{x} - u(x,\theta, t)) dt,
\end{equation}
so that 
\begin{widetext}
\begin{equation}
\begin{aligned}
\diff{\tilde{\mathcal{L}}(x(T))}{\theta}{} &=  \diff{\mathcal{L}(x(T))}{x(T)}{} \diff{x(T)}{\theta}{} - \mathfrak{A}(T) \diff{x(T)}{\theta}{} - \diff{}{\theta}{} \int_0^T \dot{\mathfrak{A}}(t) (x(t) - u(x,\theta, t)) dt \\
&=  \diff{\mathcal{L}(x(T))}{x(T)}{} \diff{x(T)}{\theta}{} - \mathfrak{A}(T) \diff{x(T)}{\theta}{} + \int_0^T \dot{\mathfrak{A}}(t) \diff{x(t)}{\theta}{} + \mathfrak{A}(t) \diff{u(x,\theta, t)}{x(t)}{}\diff{x(t)}{\theta}{} + \mathfrak{A}(t)\diff{u(x,\theta, t)}{\theta}{} dt.
\end{aligned}
\end{equation}
\end{widetext}
From this result, we then choose $\mathfrak{A}$ so that
\begin{equation}
\dot{\mathfrak{A}}(t) = -\mathfrak{A}(t) \diff{u(x,\theta, t)}{x(t)}{},\qquad \mathfrak{A}(T) = \diff{\mathcal{L}(x(T))}{x(T)}{}
\end{equation}
in order to write the gradient as
\begin{equation}
\diff{\mathcal{L}(x(T))}{\theta}{} = -\int_T^0 \mathfrak{A}(t)\diff{u(x,\theta, t)}{\theta}{} dt,
\end{equation}
which is solved backward in time because we know the final condition for the adjoint $\mathfrak{A}(T)$.

The stochastic variant of this algorithm is operationally similar to the procedure outlined above and is particularly straightforward for Stratonovich SDEs (the convention we use in numerical experiments with current-like observables)~\cite{li_scalable_2020}.

\twocolumngrid 
\bibliography{refs,hugo}
\end{document}

%% file: preamble.tex
\usepackage[x11names, rgb, html]{xcolor}

\definecolor{red}{HTML}{f54b1a}
\definecolor{pink}{HTML}{d19eb1}
\definecolor{orange}{HTML}{d3772e}
\definecolor{yellow}{HTML}{ebe85d}
\definecolor{green}{HTML}{0f6852}
\definecolor{lightblue}{HTML}{01abe9}
\definecolor{darkblue}{HTML}{1b346c}
\definecolor{tan}{HTML}{e5c39e}
\definecolor{darktan}{HTML}{af9e73}
\definecolor{grey}{HTML}{c3ced0}
\definecolor{darkgrey}{HTML}{9dadc4}
\definecolor{black}{HTML}{110d1b}
\definecolor{white}{HTML}{f1f8f1}

\usepackage{hyperref}
\hypersetup{
  colorlinks = true,
  linkcolor = red,
  citecolor = darkblue,
  urlcolor = lightblue
}

\usepackage{graphicx}

\usepackage{amsfonts,amssymb, amsmath}
\usepackage{float}
\makeatletter
\let\newfloat\newfloat@ltx
\makeatother

\usepackage{algorithm}
\usepackage[noend]{algpseudocode}
\algrenewcommand{\algorithmiccomment}[1]{$\vartriangleright$ #1}
\algrenewcommand{\algorithmicreturn}{\textbf{Return: }}
\algnewcommand\algorithmicinput{\textbf{Input: }}
\algnewcommand\Input{\State \algorithmicinput}

\newcommand{\diff}[2]{\frac{\partial#1}{\partial #2}}
\newcommand{\Diff}[3]{\frac{d^{#3}#1}{d #2^{#3}}}

\def\thetab{\boldsymbol{\theta}}

\def\bb{\boldsymbol{b}}

\def\xb{\boldsymbol{x}}

\def\Xb{\boldsymbol{X}}
\def\Xbu{\boldsymbol{X}^u}
\def\Wb{\boldsymbol{W}}

\def\grad{\nabla}

\def\RR{\mathbb{R}} \def\NN{\mathbb{N}} 
\def\EE{\mathbb{E}}\def\PP{\mathbb{P}}

\def\Var{\mathrm{Var}}

\def\<{\langle} \def\>{\rangle}